\documentclass[usenatbib]{mn2e}
\usepackage{epsfig}
\usepackage{amsmath,amssymb}
\newcommand{\be}{\begin{equation}}
\newcommand{\ee}{\end{equation}}

\newcommand{\dd}{{\rm d}}

\newcommand {\rhos} {\rho_{\rm s}}
\newcommand {\Rhos} {$\rhos$\ }
\newcommand {\rs} {r_{\rm s}}
\newcommand {\Rs} {$\rs$\ }

\newcommand {\zic} {z_{\rm IC}}

\newcommand {\msim}   {M_{\rm SIM}}
\newcommand {\mnbody} {M_{\rm Nbody}}

\newcommand{\hMsol}{{\,\rm h^{-1}M}_\odot}
\newcommand{\kpc}{{\,\rm kpc}}
\newcommand{\Mpc}{{\,\rm Mpc}}
\newcommand{\hMpc}{\,h^{-1}\Mpc}
\newcommand{\hkpc}{\,h^{-1}\kpc}

\title{Secondary infall and dark matter haloes} 
\author[Ascasibar et al.]
{
  Yago Ascasibar$^1$\thanks{E-mail: yago@aip.de},
  Yehuda Hoffman$^2$
  and
  Stefan Gottl\"ober$^1$
  \\
  $^1$Astrophysikalisches Institut Potsdam,
  An der Sternwarte 16, Potsdam 14482, Germany
  \\
  $^2$Racah Institute of Physics,
  Hebrew University, Jerusalem 91904, Israel
}
\date{\today}

\begin{document}

\maketitle

\begin{abstract}
We test the Secondary Infall Model (SIM) by direct comparison with the results of N-body simulations.
Eight cluster-size and six galactic-size dark matter haloes have been selected at $z=0$ and re-simulated with high resolution.
Based on their density profiles at the initial redshift, we compute their evolution by the SIM, assuming a simple prescription for the angular momentum.
A comparison of the density profiles obtained by the SIM and the numerical experiments at $z=5$, $1$ and $0$ shows that, for most of the haloes at most epochs, the SIM reproduces the simulated mater distribution with a typical fractional deviation of less than $40$ per cent over more than six order of magnitudes in the density.
It is also found that, within the SIM framework, most of the diversity in the shape of the density profiles at $z=0$ arises from the scatter in the primordial initial conditions rather than the scatter in the angular momentum distribution.
A crude optimization shows that a similar degree of agreement is obtained for galactic and cluster haloes, but the former seem to require slightly higher amounts of angular momentum.
Our main conclusion is that the SIM provides a viable dynamical model for predicting the structure and evolution of the density profile of dark matter haloes. 
\end{abstract}


%%--------------------
\section{Introduction}
\label{sec:intr}
%%--------------------

In the hierarchical clustering paradigm, cosmological structure evolves from a primordial density perturbation field via gravitational instability.
One of the basic outcomes of the process is the formation of virialized objects, usually referred to as dark matter (DM) haloes.
The problem of the collapse and virialization of DM haloes can be easily formulated, as it depends on a single, relatively simple, physical process: the dissipationless gravitational interaction of a system composed of a large number of point-like particles.

Yet, the long range and non-linear nature of the gravitational N-body problem has made it extremely elusive to rigorous analytical treatment.
On a phenomenological level, however, tremendous progress has been achieved by means of numerical experiments.
After roughly three decades of N-body simulations \citep[starting with][among others]{Aarseth+79,vanAlbadavanGorkom77,White78}, this technique has reached the level at which different  algorithms and numerical codes have converged to yield an overall consistent description of the formation, evolution and internal structure of DM haloes \citep{Knebe+00,Knebe+01_MLAPM}.

A major breakthrough was the suggestion by \citet{NFW96,NFW97} (hereafter NFW) that the density profile of simulated DM haloes can be fitted by a simple analytical function,
\begin{equation}
\label{eq:nfw}
\rho(r) = \frac{4\rhos}{(r / \rs)( 1 + r / \rs)^2},
\end{equation}
where the two free parameters \Rhos and \Rs represent a characteristic density and radius of the halo.
There is general consensus within the N-body community that the so-called NFW profile is able to provide a reasonably good fit to the numerical results in a variety of cold dark matter (CDM) cosmologies, although some doubts have been cast on the exact value of the logarithmic slope at the centre \citep[the so-called density cusp, see e.g.][]{Moore+98,Moore+99,Ghigna+98,Ghigna+00,FukushigeMakino97,FukushigeMakino01,Fukushige+04,Power+03,Hayashi+04,Navarro+04}, as well as on the degree of `universality' of the fit, i.e. its dependence on the underlying cosmological model, the mass accretion history of the halo, or its environment \citep[e.g.][]{JingSuto00,Klypin+01,Ricotti03,AvilaReese+05,Maulbetsch+_06}.

Observationally, the DM density profile must be indirectly inferred.
In rich clusters of galaxies, estimates based on X-rays \citep[e.g.][and references therein]{VoigtFabian06} or gravitational lensing \citep[e.g.][]{Dahle03,Gavazzi+03,BartelmannMeneghetti04,DalalKeeton_03} seem to indicate that the mass distribution is indeed well described by equation~(\ref{eq:nfw}).
On galactic scales, observations are much more difficult to interpret.
Dwarf spiral and low surface brightness galaxies are most likely dominated by DM in their centres, but the mass distribution inferred from rotation curves suggests a constant DM density core rather than a steep cusp  \citep[e.g.][]{FloresPrimack94,Moore94}.
Recent analyses show that observational data may actually be consistent with steeper profiles once the effects of inclination, non-circular orbits and triaxiality of the dark matter haloes are accounted for \citep{Hayashi+04,HayashiNavarro_06}, but the controversy is still unresolved \citep[e.g.][]{Gentile+04,deBlok05}.

It is therefore not clear whether the existence of a more or less `universal' density profile is supported by observational data, nor whether the possible disagreement is due to our lack of an understanding of the very complicated process of galaxy formation or it calls for an overall revision of the the standard model of cosmology.
In any case, the qualitative shape of the density profile (shallower than isothermal in the inner parts and steeper in the outer), as well as (to some extent) its `universality' constitute a very robust prediction of the currently accepted DM structure formation paradigm.

Unfortunately, the NFW  fitting formula provides a good phenomenological description of the density profile of simulated DM haloes, but it does not provide a physical understanding of its origin.
Ideally, one would have liked to have a self-consistent analytical model that enabled the calculation, from first principles only, of the density profile expected for a given cosmology, halo mass, formation time, environment, or whatever physically meaningful parameter that is found to play a relevant role.

No such model exists as yet, but we are now close to having one.
Much of the analytical work on the formation of DM haloes is based on the seminal paper of \citet{GunnGott72} on the dissipationless collapse of a spherical homogeneous perturbation in an otherwise expanding Friedmann universe.
This was followed by \citet{Gunn77}, who considered the collapse of a spherical inhomogeneous perturbation and the role of shell crossing.
It also studied the case of the secondary infall, namely the late infall of shells onto an already collapsed and virialized perturbation, and showed that a self-similar solution can be found upon the use of adiabatic invariance.

\citet{Gunn77} was followed by two different lines of research.
On one hand, \citet{FillmoreGoldreich84} and \citet{Bertschinger85} independently found self-similar solutions of the collapse of scale-free spherical perturbations in an Einstein-de Sitter universe.
On the other hand, \citet{HoffmanShaham85} analyzed the dependence of the structure of proto-haloes on the primordial power spectrum and the way it affects the density profile of virialized DM haloes.

The basic predictions of the secondary infall model (SIM) were qualitatively confirmed by N-body simulations \citep[e.g.][]{Quinn+86,Efstathiou+88,Crone+94}, which prompted further study and extension of the SIM, mostly focusing on the nature of the initial conditions, on the improvement of the dynamical model and on its cosmological implications \citep[e.g.][]{RydenGunn87,Hoffman88,Ryden88_j,Ryden88_p,ZaroubiHoffman93,LokasHoffman00}.
A common assumption made in these studies was that the halo particles are moving along radial orbits, and the generic result that emerges is that the inner density profile is roughly given by $\rho \propto r^{-2}$.
Thus, a cuspy density profile is a generic outcome of the SIM.
Actually, the predicted logarithmic slope is even steeper than the one measured in N-body experiments.

The next major improvement of the SIM was the introduction of non-radial motions, namely a distribution of angular momentum of individual particles, not necessarily implying, on average, a total angular momentum of the halo.
A number of authors \citep[e.g.][]{WhiteZaritsky92,Ryden93,AvilaReese+98} have pointed out that non-radial motions flatten the inner density profile.
\citet{Nusser01} extended the self-similar solutions to include a distribution of angular momentum, and several recent studies show that, by introducing enough angular momentum, the density profile predicted by the SIM can have a $\rho \propto r^{-1}$ density cusp \citep{Hiotelis02,leDelliouHenriksen03,Ascasibar+04,Williams+04,Lu+06}.

Within the SIM framework, there are two key ingredients that determine the shape of the virialized density profile: the initial density profile of the proto-halo and the angular momentum distribution.
The primordial density has been determined either by employing the statistical properties of Gaussian random fields \citep{HoffmanShaham85,BBKS86,Hoffman88,Ascasibar+04} or by means of the mass accretion history of DM haloes \citep{AvilaReese+98,NusserSheth99,Lu+06}.
The angular momentum distribution has always been considered to be a free parameter, determined so as to yield best agreement with simulations.
\citet{Ascasibar+04} pushed the study of the SIM one step further by making a detailed comparison between the density profile predicted by the model with the actual numerically simulated DM haloes rather than the NFW fit.
These authors assumed the initial density profile to be given by the average expectation around a local maximum of the primordial perturbation field, determined by the peak height and its smoothing scale.
Optimizing over these two parameters and the angular momentum distribution, the resulting density profiles were in close agreement with the simulated ones, with an accuracy comparable to that of the best-fitting NFW profile.

The aim of the present paper is to go one step beyond \citet{Ascasibar+04} in order to assess the validity of the SIM as a tool to understand the physical origin of the DM density profile.
More precisely, we attempt to recover the density profile of actual simulated haloes, but instead of assuming a functional form for the initial conditions, we identify the primordial peak in the initial snapshot of the simulation and use the density profile around that point as an input for the SIM.
We calculate the entire dynamical evolution of the halo upon assuming only one free parameter (the angular momentum distribution of the DM particles), and compare the density profile thus obtained with the full N-body simulation.
We consider this as the ultimate test of the validity of the SIM, at least in the framework of the CDM-like cosmology.

The paper is structured as follows:
The basic principles of the SIM are reviewed in \S~\ref{sec:sim}.
Numerical simulations are described in \S~\ref{sec:nbody}, and the comparison of the SIM-calculated density profiles with the results of the full N-body simulation is presented in \S~\ref{sec:results}.
A discussion and summary of the results is given in \S~\ref{sec:discu}.
Technical details of the SIM and its implementation are given in Appendix~\ref{sec:appSIM}.

%%
%%------------------
\section{Secondary Infall Model}
\label{sec:sim}
%%------------------

It is well known from cosmological N-body simulations that halo formation proceeds through a series of violent merger events involving smaller substructures.
Indeed, major mergers seem to play an important role in shaping the structure of DM haloes \citep[see e.g.][]{SyerWhite98,SalvadorSole+98,Manrique+03,RomanoDiaz+06}, and therefore one would naively expect any model based on spherical symmetry to be hopelessly irrelevant and inadequate.

The SIM is not only based on the assumption of spherical symmetry.
It further assumes that the complicated process of shell crossing, in which particles, represented by spherical shells, do not conserve their individual energies, admits a simple adiabatic invariant.
This allows one to bypass the need for a full self-consistent, one-dimensional, spherically symmetric, non-linear calculation of the dynamics of the collapsing shells, and use a semi-analytical method to calculate the equilibrium structure of DM haloes \citep[for an interesting alternative approach, see the recent work by][]{SanchezConde+_06}.
Adiabatic invariance was introduced by \citet{Gunn77}, who applied it to the case of self-similar collapse and this was  extended  to the general non-power law case by \citet{ZaroubiHoffman93}. Angular momentum was  first included by \citet{RydenGunn87}, and then in a more rigorous way by \citet{Nusser01}.

A very brief description of the SIM is given here and a detailed account of the SIM dynamical model and its implementation is given in Appendix~\ref{sec:appSIM}.
The physical model that underlies the SIM is based on an assumed primordial density profile of the proto-object.
 Given that, the trajectory of a given mass shell is followed analytically until it reaches its turn-around radius.
The shell does not experience shell crossing before it turns around, its energy is conserved and the calculation is exact.
From that point, one needs to resort to the assumption that the shell trajectory admits an adiabatic invariant, whereby the the maximum radius reached by the shell times the enclosed mass is approximately conserved.
The transformation from the shell trajectories to a density profile is done by calculating the amount of time a given shell spends within a given radial interval.
The angular momentum of each shell is determined by a free parameter, $\eta$, that measures the square of the angular momentum of dark matter particles in terms of the maximal possible value and controls both the minimum radius reached by the shell and its time-radius relation.
The physical basis for invoking adiabatic invariance is that, as a given shell turns around and settles into equilibrium, it constitutes only a small perturbation to shells that have already collapsed and virialized.

It follows that the evolution of the resulting DM halo is completely determined by its initial density profile and the intrinsic angular momentum distribution.
In the absence of a detailed information on the primordial structure of a given halo, one must resort to statistical means.
Two main approaches have been used here.
One relies on the assumption that  DM haloes are seeded by local density maxima of the primordial perturbation field, which is assumed to be Gaussian.
The mean density profile around a local density maximum can then be readily calculated \citep{BBKS86} and used to set the initial density profile of the proto-halo \citep{HoffmanShaham85,Hoffman88,Ascasibar+04}.
Alternatively, one can use the mass accretion history of haloes, together with the spherical top-hat model, to compute the mean density profile of proto-haloes \citep{NusserSheth99,AvilaReese+98,Lu+06}. 

The other key element of the SIM is the angular momentum distribution of the DM particles. The amount of time a particle spends near the centre is controlled by its angular momentum, and thereby this quantity affects the `cross-talk' and energy exchange between the shells.
For pure radial orbits, the \citet{FillmoreGoldreich84} self-similar solution is recovered almost independently of the initial conditions, and an $r^{-2}$ density cusp is obtained  both numerically \citep{Huss+99} and analytically \citep{LokasHoffman00}.
For the case of circular orbits, the turn around radius density profile is recovered \citep{HoffmanShaham85}.

The methodology of the present paper is to  calculate the evolution of the density profile of individual DM haloes by the SIM and compare it with the simulated profiles.
The initial conditions of the haloes are provided by the simulations themselves, rather than by some general considerations.
Namely, the simulations are playing here a dual role, as they calculate the evolution of the selected objects, but also provide a mapping from the final virialized structures to their initial conditions.
Here we improve on earlier studies of the SIM by testing the time evolution of the haloes and by replacing an assumed parametric fit to the initial conditions with the actual ones.

As in previous work, we simply assume an angular momentum distribution, and the empirical relation (\ref{eq:eta}) has been invoked between the specific angular momentum of DM particles, the mass enclosed by the corresponding spherical shell and its turn-around radius.
The parameter $\eta$ controls the amount of angular momentum injected, with the extreme values $\eta=0$ and $\eta=1$ corresponding to purely radial and circular orbits, respectively (see Appendix~\ref{sec:appSIM} for details).

%%
%%------------------
\section{N-body simulations}
\label{sec:nbody}
%%------------------

We use the Adaptive Refinement Tree code \citep{Kravtsov+97_ART} to follow
the evolution of structure in the concordance $\Lambda$CDM cosmology
($\Omega_m = 0.3$, $\Omega_{\Lambda} = 0.7$, $h=0.7$, $\sigma_8
= 0.9$).

We are interested in the evolution of cluster-sized as well as galaxy-sized haloes.
To construct suitable initial conditions, we use the
multiple-mass technique \citep{Klypin+01}. In a first step we created
an unconstrained random realization on an $N=1024^3$ or $N=2048^3$ grid.
The initial displacements and velocities of the particles
were calculated using all waves ranging from the fundamental mode
$k=2\pi/L$ to the Nyquist frequency $k_{\rm ny}=2\pi/L\times
N^{1/3}/2$.  Initial conditions at lower resolution were produced by
merging those small-mass particles to get in total $128^3$ particles
and assigning to merged particles a velocity and a displacement of one
of the small-mass particles.  The whole box of $80\hMpc$ size
was first simulated at this low resolution, starting at redshift
$\zic=50$.

Eight clusters have been selected from this low-resolution simulation,
and the multiple-mass technique was used to set up high-resolution
initial conditions.  Namely, a Lagrangian region corresponding to a
sphere of radius equal to two virial radii around each halo at $z=0$
was sampled with particles of mass $m_{\rm p}=3.16\times 10^8\hMsol$,
corresponding to an effective number of $512^3$ particles in the box.
The high mass-resolution region was surrounded by layers of particles
of increasing mass.  The force resolution reached is $2.4\hkpc$ (two
times the size of the highest refinement level cell).

To study the galaxy sized haloes, we have selected a filamentary region
within a box of $80\hMpc$ size. With the same resimulation technique
as described above, this region has been simulated with 150 million
particles, which corresponds to an effective number of $2048^3$
particles. Thus, the mass resolution is $5.0 \times 10^6\hMsol$, and
force resolution reaches $0.6\hkpc$. Six galactic-scale haloes have been selected.

For each object, we find the density maximum at $z=0$ by iteratively computing the centre of mass, starting with a reasonable initial guess and a search radius of $0.1\hMpc$.
Once convergence is reached, the search radius is reduced by ten percent, and the process is repeated until the sphere encloses less than $10^3$ particles.

Then, the positions of the $10^4$ particles closest to the density maximum at $z=0$ are traced back to the initial conditions, and we use only those positions to locate the primordial peak at $z=\zic$.
It is interesting to note that, in many cases, the particles near the centre at $z=0$ belong to different objects at $\zic$, which merged at some point during the history of the halo.
Our procedure identifies the density peak corresponding to the most massive progenitor, which is a well defined entity, even at such a high redshift.
Once the peak is located, we recompute its position, this time taking into account all particles at $\zic$.
The difference in position is usually not very large, but it has a significant impact on the shape of the primordial density profile near the centre.

Finally, we trace the $10^4$ particles closest to the density maximum at $\zic$ and locate the descendants at $z=5$, $1$ and $0$.
We find the density maximum at those redshifts, taking into account all particles in the corresponding snapshots, and compute the simulated density profiles.

%%
%%-----------------
\section{Results}
\label{sec:results}
%%-----------------
%%

The dynamical evolution of the selected haloes has been calculated by both the SIM and the N-body simulation.
By construction, the initial profiles coincide at $\zic$, and we compare the results obtained by both methods at different redshifts (namely $z=5$, $1$ and $0$).
As explained in Section~\ref{sec:sim} and Appendix~\ref{sec:appSIM}, angular momentum of the spherical shells in the SIM is determined by the $\eta$ parameter.
For each object, the SIM has been applied with a range of angular momentum values given by $\eta=0.10, 0.11, 0.12, ..., 0.22$.

\begin{figure*}
\epsfig{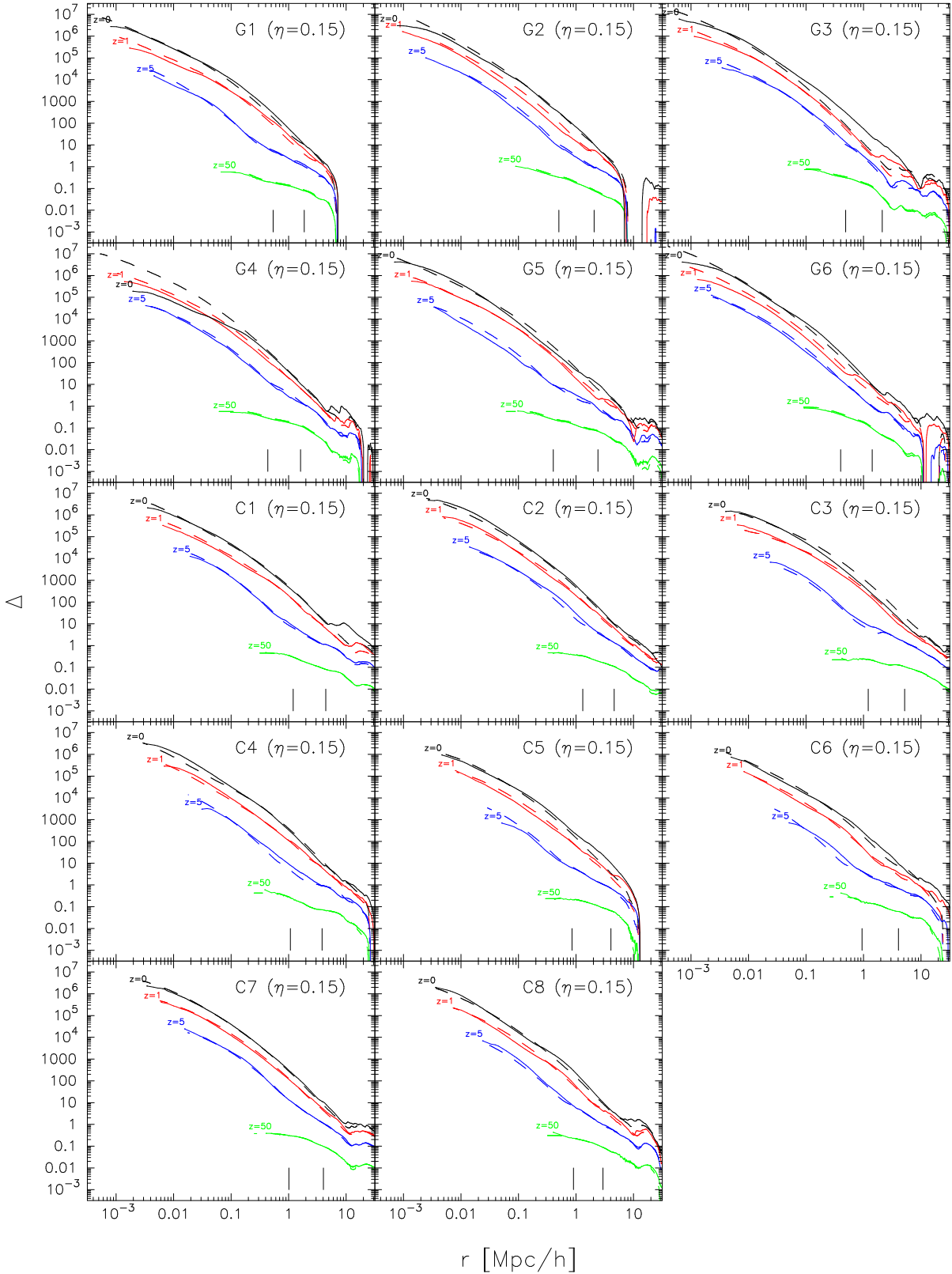}
\caption{\label{fig:prof0.15}
Cumulative overdensity profiles, $\Delta(r)\equiv \frac{3M(r)}{4\pi\Omega_m\rho_c r^3} -1$, of our six galactic haloes (designated G1 to G6) and eight clusters (C1 to C8).
Dashed and solid lines display the results of the SIM and the N-body experiments, respectively.
From top to bottom, they correspond to $z=0$, $1$, $5$, and $50$ (the initial time of the simulations).
The SIM is applied with the same angular momentum distribution for all haloes, given by $\eta=0.15$.
The location of the virial (left) and turnaround (right) radii at the present epoch is indicated by the small vertical lines near the $x$-axis.
}
\end{figure*}

A first comparison of the SIM and N-body profiles is presented in Figure~\ref{fig:prof0.15}, where all the SIM density profiles have been computed for $\eta=0.15$.
When this parameter is allowed to vary from object to object, a slightly better agreement can be obtained in the inner regions.
However, it is not our aim to \emph{fit} the N-body data, but to \emph{predict} the evolution of the density profile, using only information at $\zic$.
Therefore, no attempt has been made here to make a quantitative determination of best-fitting $\eta$ for every object.

\begin{figure*}
\epsfig{file=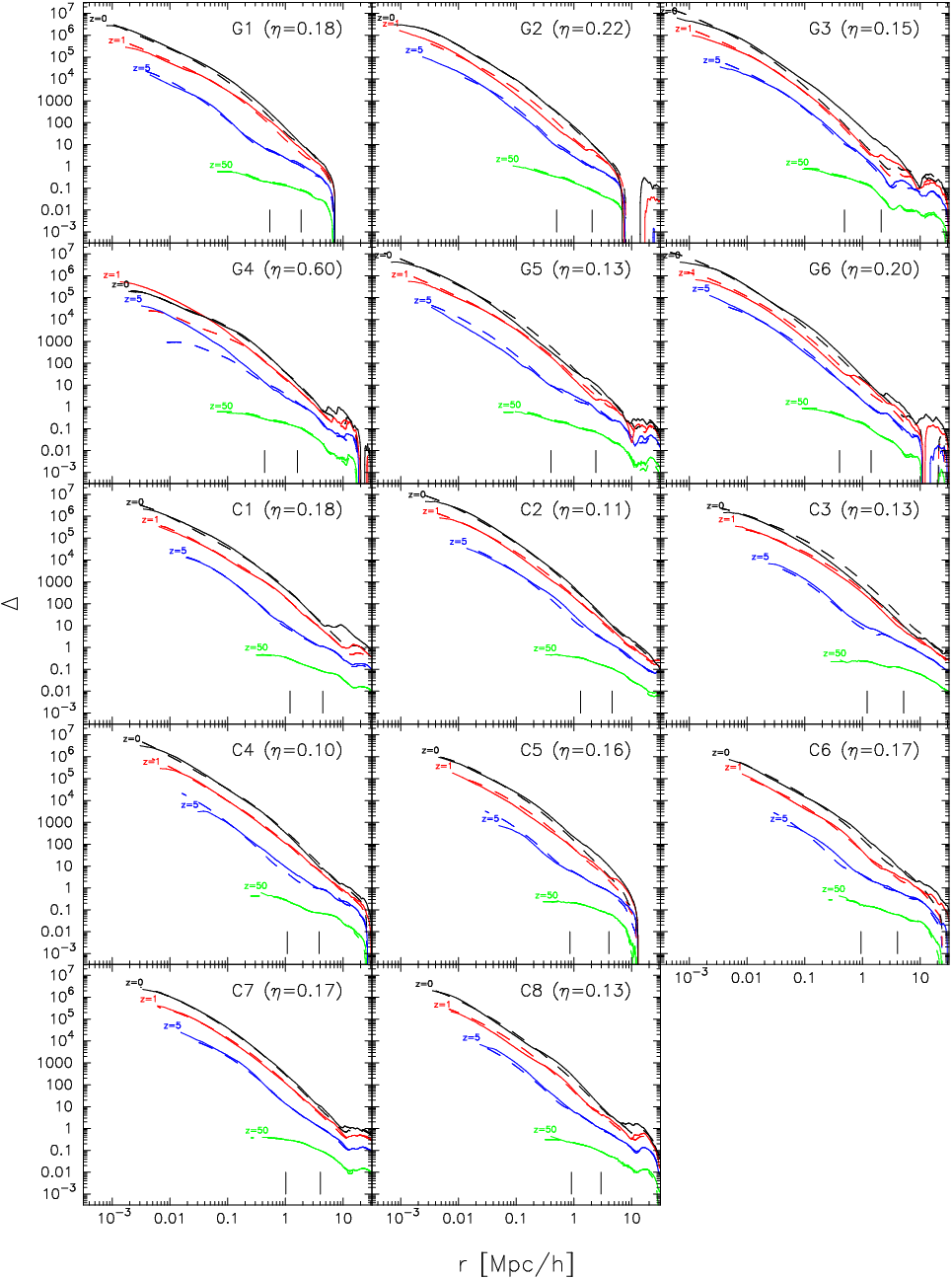,width=0.9\hsize}
\caption{\label{fig:profeta}
Same as Figure~\ref{fig:prof0.15}, but with the optimal value of $\eta$ for each cluster, as indicated in the individual frames.
}
\end{figure*}

A qualitative estimate may nevertheless shed some light on the most plausible values of $\eta$, the associated scatter, and perhaps the dependence on mass, environment, or other factors.
Figure~\ref{fig:profeta} shows again the comparison between SIM and N-body profiles, but in this case the value of $\eta$ has been individually optimized by visual inspection.
In some cases, it might be possible to obtain an even closer correspondence between both methods if we also let $\eta$ vary with time, but in general terms, our results suggest that the introduction of such an additional degree of freedom is not necessary.

It is interesting to compare here the galactic- and cluster-scale haloes. The size of the two samples is very small (six and eight objects) and no attempt  is made here for any formal statistics, yet we find that in general galactic haloes are better fitted by higher values of $\eta$. The G4 halo constitutes an enigmatic case. At z=0, the SIM provides an excellent fit to the simulated profile for an extremely high value of $\eta=0.60$. A close inspection of the halo reveals that it is in the process of a major merger, and yet the agreement at $z=0$ is remarkably good.

\begin{figure*}
\epsfig{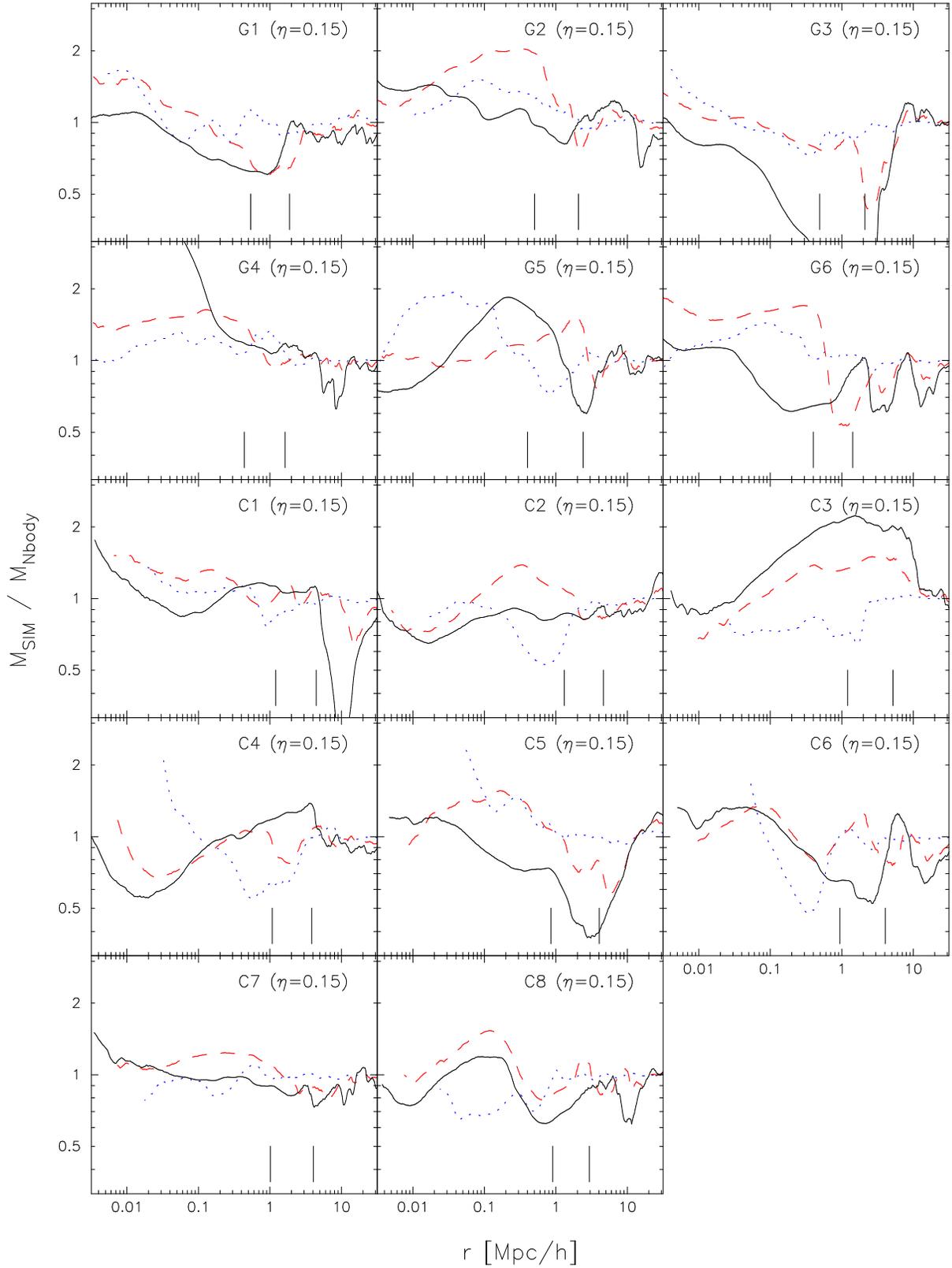}
\caption{\label{fig:frac0.15}
Ratio between the cumulative mass profiles, $M(r)$, calculated by the SIM and the N-body experiments, plotted for $z=5$ (dotted lines), $1$ (dashed lines) and $0$ (solid lines).
The SIM is applied with the same angular momentum distribution for all haloes, given by $\eta=0.15$.
The location of the virial (left) and turnaround (right) radii at the present epoch is indicated by the small vertical lines near the $x$-axis.
}
\end{figure*}
\begin{figure*}
\epsfig{file=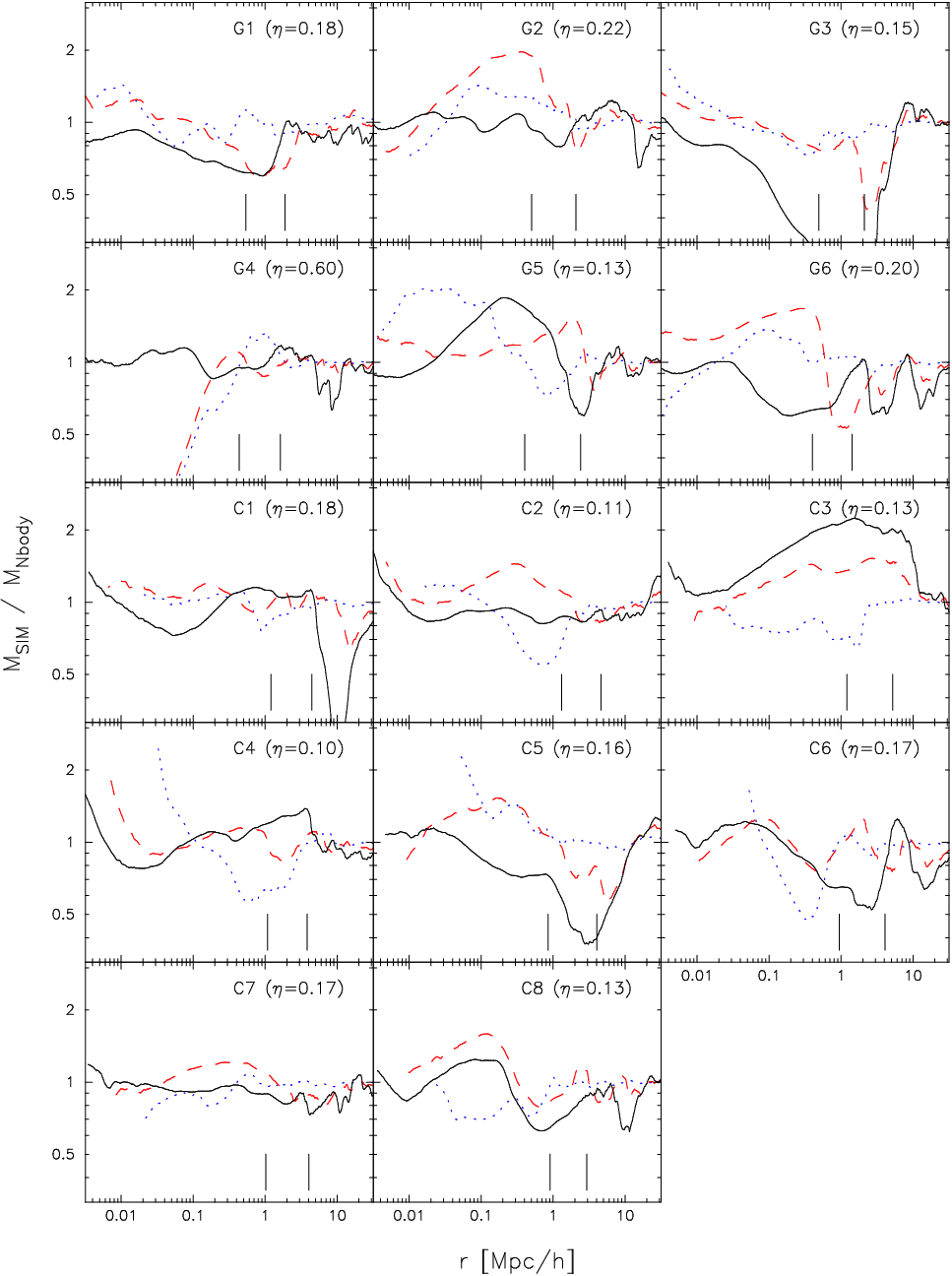,width=0.9\hsize}
\caption{\label{fig:fraceta}
Same as Figure~\ref{fig:frac0.15}, but with the optimal value of $\eta$ for each cluster, as indicated in the individual frames.
}
\end{figure*}

\begin{figure*}
\epsfig{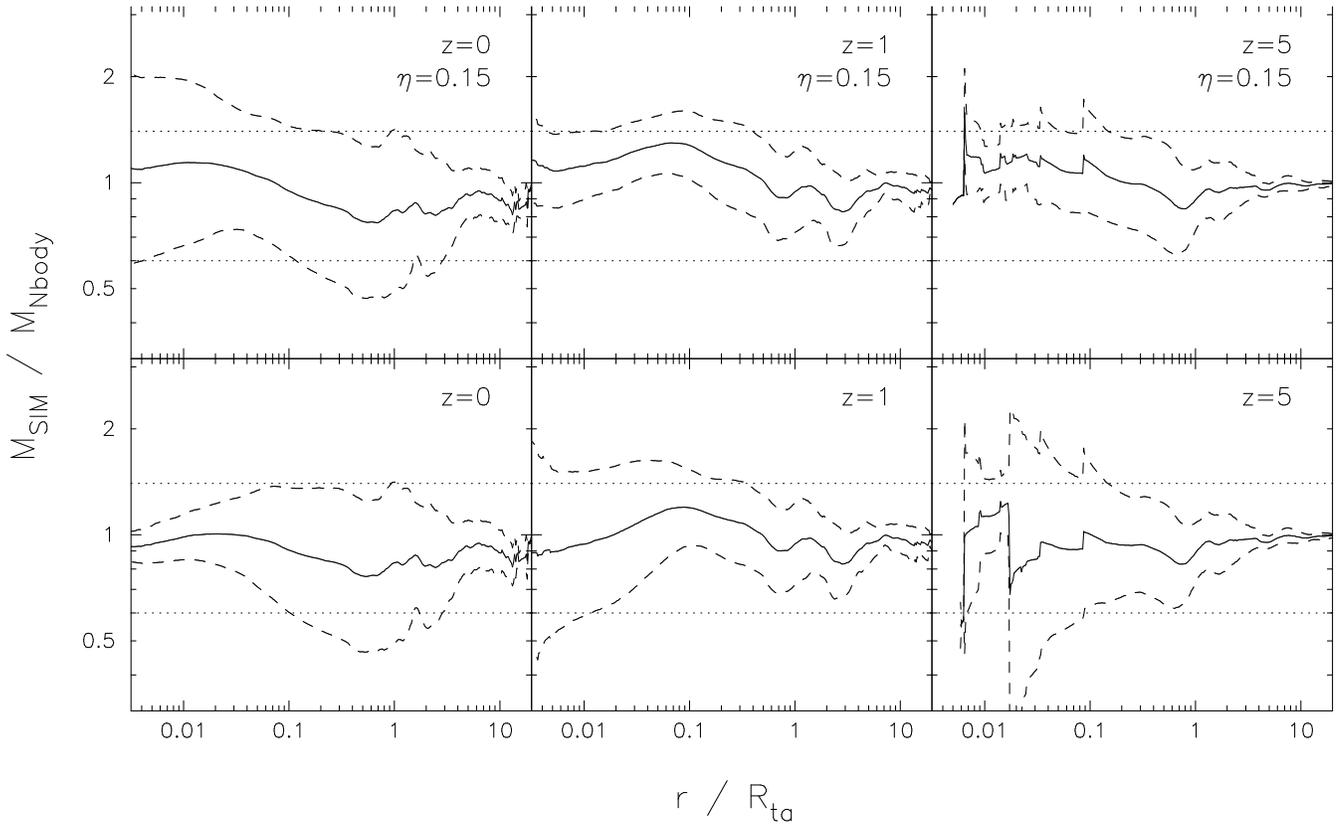}
\caption{\label{fig:dis}
Mean (solid line) and standard deviation (dashed lines) of the ratio between the enclosed mass predicted by the SIM and that computed by the N-body simulation, averaged over all (galactic- and cluster-size) haloes, scaled by their turn-around radius.
Results obtained for fixed ($\eta=0.15$) and individually-optimized $\eta$ are plotted in the upper and lower panels, respectively.
The three columns correspond to redshifts $z=0$, $1$ and $5$.
Horizontal dotted lines indicate a fractional deviation of $\pm 40$ per cent.
}
\end{figure*}

We plot in Figure~\ref{fig:frac0.15} (for a fixed $\eta=0.15$) and Figure~\ref{fig:fraceta} (for individually optimized values) the ratio between the enclosed mass profile predicted by the SIM and the one calculated by the N-body simulation.
The mean and the standard deviation of $\msim/\mnbody$ are shown in Figure~\ref{fig:dis}, where the statistics is calculated over all haloes, galactic and cluster size, normalizing the radius by the turn-around radius of each object.
The SIM calculations are done for a fixed $\eta$ (upper row) and for optimized $\eta$ per individual haloes (lower row).
In both cases, the plots show very little bias ($\sim20$ per cent) and a scatter not larger than $40$ per cent over most of the radial range.

\begin{figure*}
\epsfig{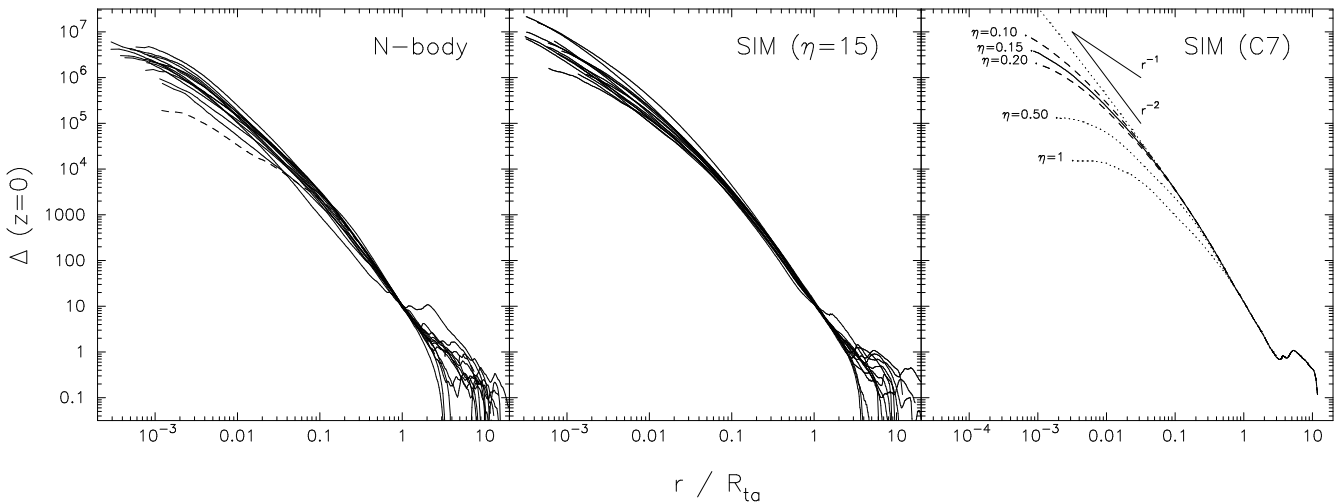}
\caption{\label{fig:div}
Diversity in the shape of the density profile.
Left panel: radial density profile, scaled by the turn-around radius, of all simulated objects at the present epoch.
Middle panel: profiles calculated by the SIM, assuming a fixed angular momentum distribution given by $\eta=0.15$.
Right panel: Effect of angular momentum in the SIM.
Taking object C7 as reference, the solid ($\eta=0.15$) and two dashed lines ($\eta=0.10, 0.20$) encompass the typical values of $\eta$ found in our study.
The extreme cases $\eta=0$, $0.5$, and $1.0$ (dotted lines) and power laws $r^{-1}$ and $r^{-2}$ (small solid lines) are also shown for reference.
The comparison shows that, within a typical range of $\eta$, the diversity in the shapes of the density profile at $z=0$ is dominated by variations in the primordial structure rather than by the amount of angular momentum of each halo.
}
\end{figure*}

Finally, we are also interested in understanding the physical origin of the diversity of density profiles found in the N-body experiments.
In terms of the SIM formalism, we would like to know if this diversity arises from scatter in the primordial conditions, in the specific angular momentum distribution of each halo, or both.
The range of profiles exhibited by the simulated objects at $z=0$ is shown on the leftmost panel of Figure~\ref{fig:div}.
The middle panel shows how the initial conditions  are affecting  the shape of the density profiles by grouping together all the SIM-calculated density profiles, evaluated at $z=0$, for a fixed $\eta=0.15$.
The role of the angular momentum is illustrated on the rightmost panel, where the density profile predicted by the SIM for cluster C7 at $z=0$ is plotted for $\eta=0.10$, $0.15$ and $0.20$.
The cases $\eta=0$ (radial orbits, expected to yield an $r^{-2}$ cusp), $\eta=1/2$, and $\eta=1$ (circular orbits, reflecting the turnaround density profile, and hence expected to result in a core structure) are given for reference.  
From Figure~\ref{fig:div}, we conclude that most of the diversity in the shapes of the virialized DM haloes is contributed by the variability in the initial conditions.
The distribution of angular momentum, at least within the range of $\eta=0.15 \pm 0.05$ that seems to describe our sample, plays only a secondary role.

%%
%%------------------
\section{Discussion}
\label{sec:discu}
%%------------------
%%

%general summary
The main result of the present work is that the secondary infall model provides a valid theoretical framework for calculating the structure and evolution of DM haloes in an expanding universe, and that its predictions with respect to the density profiles are in close agreement with full N-body simulations.
Comparing the SIM and simulated cumulative density profiles over more than six orders of magnitude, we find the typical discrepancy to be better than $40$ per cent.
This level of agreement extends to the time evolution of the density profile and it is not limited to the final time snapshot only.
Within the SIM framework, most of the diversity in the density profiles of DM haloes is contributed by the scatter in the primordial profiles rather than the scatter in the angular momentum.

%Cluster vs galactic haloes
Galactic-size haloes are less isolated objects than cluster haloes, in the sense that they have more similar-mass companions.
The simulated galaxies have been selected from a typical filamentary region and naive reasoning would lead one to expect these objects to be less suitable for the application of the SIM compared with clusters. This is not the case. The agreement between the N-body and SIM density profile of galactic haloes is roughly as good as for clusters.
The only difference is that the crude optimization over the angular momentum parameter, $\eta$, finds that galaxies require slightly higher values of $\eta$, consistent with the fact that the galactic haloes are more torqued and hence should have acquired more angular momentum. This in turn gives further support to the SIM in the sense that the value of $\eta$ is not a mere free parameter but is related to the actual angular momentum of haloes.

%G4 halo
The galactic-scale halo G4 deserves a special attention. The SIM-predicted profile with $\eta=0.15$ provides a very bad fit to the simulated density profile at $z=0$. Yet, for the very high value of $\eta=0.60$, a very good fit is obtained for the present epoch.
Such a high value constitutes a remarkable exception, and no other object has a best-fitting $\eta$ anywhere close to $0.60$.
Visual inspection of the G4 halo reveals a major merger in progress at $z=0$, and the halo is clearly not in virial equilibrium.
In spite of its dynamical state, the density profile of the halo is extremely well fitted by the SIM.
Is such an agreement a mere coincidence, or does the SIM actually reproduce the internal structure the system? This intriguing question has led us to further investigate the structure of the G4 halo, and we found that it actually displays an extremely high amount of angular momentum due to the merger event.

%Possible extensions:  velocity structure, anisotropy of orbits, phase space density, entropy
Although the present work focuses on the density profile, it would be desirable to extend the formalism outlined in Appendix~\ref{sec:appSIM} so as to compute their full dynamical structure of DM haloes; more precisely, the radial and tangential components of the velocity dispersion.
Derived quantities, such as the angular momentum distribution or the anisotropy profile, can be readily calculated, and their dependence on the assumed value of $\eta$ would be extremely helpful in validating our simple prescription for angular momentum quantitatively.

Moreover, one could use the upgraded formalism to address additional questions raised by numerical N-body simulations, such as the relation between the logarithmic slope of the density profile and the anisotropy of the velocity dispersion \citep{HansenMoore06}, or the power-law radial dependence of the coarse-grained phase-space density \citep{TaylorNavarro01,Ascasibar+04,Rasia+04}. This quantity is somewhat analogous to the `entropy' of the DM, formally defined so as to mimic the ideal gas thermal entropy, and it has indeed been found that they are closely related in adiabatic gasdynamical simulations \citep{Faltenbacher_06}.

%general remarks on the SIM
Eventually, the final goal would be to achieve a totally self-consistent prescription to evaluate the angular momentum distribution of DM haloes, which of course can only be achieved through an understanding of its physical origin.
Although strict spherical symmetry would imply that the angular momentum of individual particles must be exactly conserved (and hence it should be imprinted by the initial conditions), several processes may be responsible for the conversion of radial into tangential motions in a more realistic case.

For instance, analytical and numerical studies have shown that a predominantly radial collapse is unstable to the growth of tangential motions, resulting in the so-called radial orbit instability \citep[see e.g.][and references therein]{Huss+99}, and it has been conjectured \citep{Barnes+06} that this mechanism could actually set the length scale \Rs of DM haloes, which marks the transition region from almost isotropic orbits to radial ones.

A seemingly very different point of view is provided by cosmological simulations, which are assumed to incorporate all the physics relevant for the formation of DM haloes.
Haloes in numerical experiments evolve through phases of violent `major merger' events in which objects of comparable masses merge to form a yet bigger halo.
A detailed analysis shows that the NFW-like structure, and in particular the angular momentum distribution within $\rs$, is exclusively determined by major mergers, and is hardly affected by the quiescent phases that follow \citep[][and references therein]{SalvadorSole+05,RomanoDiaz+06}.

To summarize,  two very different frames of reference for DM halo formation end up predicting very similar density profiles.
Assuming that this is not a coincidence, one should look for an underlying physical process that unites these two different approaches. 
Our results suggest that angular momentum might play an important role in that respect. 

Yet, the present study points to another key element and this is the initial configuration of the haloes.
Figure~\ref{fig:div} shows the diversity in the density profile of the virialized objects caused by the variation in their initial density structure.
It follows that the shape of the halo density profile cannot be discussed without an explicit reference to the nature of the initial conditions; a link that any complete theory of the collapse of DM haloes should incorporate.
Although a full understanding is yet to be achieved, the study of the collapse and virialization of cosmic structures that commenced with \citet{GunnGott72} has certainly provided us with many clues and a deep insight of the most relevant ingredients behind the process.

%Possible application - Tolia's idea
As a final note, we would like to suggest a potential practical application of the SIM as a complement (perhaps even substitute) of the multiple-mass technique.
High-resolution re-simulations of individual DM haloes identified in low-resolution cosmological experiments require a significant amount of CPU resources, and they can be practically performed only for a small subset of all objects.
We have shown, however, that the SIM provides a valid framework for calculating the dynamical evolution of DM haloes, and it takes about $10$~s to compute the density profile of a given object at a given epoch on a desktop PC.
This gives rise to the interesting possibility of using the SIM to insert highly-resolved spherical haloes into large-scale, low-resolution cosmological simulations.
In practice, one can use the low-resolution experiment to set up the initial conditions for all the objects and use the SIM to calculate their dynamical evolution, providing an efficient and relatively accurate way of generating extended catalogs of dynamically resolved clusters at a reasonable accuracy.

%%------------------------
\section*{Acknowledgments}
%%------------------------
Useful discussions with M. Hoeft and A. Klypin are gratefully
acknowledged.  This research has been supported by ISF-143/02 and the
Sheinborn Foundation (to YH).  YH acknowledges a Mercator
Gastprofessur at Potsdam University.  The numerical simulations
were performed at the LRZ Munich, NIC J\"ulich, and NASA Ames.

%%
%%----------------

\appendix

\section{Implementation of the Spherical Infall Model}
\label{sec:appSIM}

\subsection{Physics}

We consider the evolution of a spherical distribution $M(r)$, following the prescription described in \citet{Ascasibar+04}.
We use the initial comoving radius $x$ as a Lagrangian coordinate identifying a shell of matter enclosing a mass
\be
M_x=\frac{4\pi}{3}\Omega_m\rho_c x^3
\ee
where $\Omega_m$ denotes the current dark matter density and $\rho_c$ is the critical density.
The physical location of the shell can be written in terms of the variable $\alpha(t)$,
\be
r(t) = x a_i \alpha(t)
\ee
where $a_i=a(t_i)$ is the expansion factor of the universe at an arbitrary initial time, $t_i$.

In a homogeneous universe, $\alpha(t)$ would simply be the cosmic expansion factor, i.e. $\alpha(t)=a(t)/a_i\ \forall x$.
A spherical perturbation with overdensity
\be
\Delta_i \equiv \Delta(x,t_i) = \alpha_i^{-3} -1
\ll 1
\ee
can be obtained by slightly displacing the shells.
To first order in $\Delta_i$,
\be
r(t_i) = x a_i \alpha(t_i) \approx x a_i \left( 1 - \frac{\Delta_i}{3} \right),
\ee
while the velocity would be given by
\be
\dot r(t_i) = x a_i \dot\alpha(t_i) \approx H_i x a_i \left( 1 - \frac{2\Delta_i}{3} \right)
\ee
where $H_i\equiv \sqrt{\Omega_m a_i^{-3} + \Omega_\Lambda + (1-\Omega_m-\Omega_\Lambda) a_i^{-2}}$ corresponds to the Hubble constant at $t=t_i$.
In what follows, time will always be expressed in units of $H_i^{-1}$.

Before shell-crossing,
\be
\epsilon_1(x) =
\frac{\dot r^2}{2}-G\frac{M_x+\frac{4\pi}{3}\rho_\Lambda r^3}{r}
\ee
is a conserved quantity, analogous to the Newtonian specific energy of each shell.
With our initial conditions, $\epsilon_1(x) \approx - 5/6(H_i x a_i)^2\Delta_i$, yielding the equation of motion
\be
\dot \alpha^2=\Omega_i\alpha^{-1}+\Lambda_i \alpha^2 - K_i
\label{eqMotion1}
\ee
where $\Omega_i=\Omega_m a_i^{-3} H_0^2/H_i^2$ and $\Lambda_i=\Omega_\Lambda H_0^2/H_i^2$ are the matter and vacuum energy densities at time $t_i$, and
\be
K_i = - 1 + \Omega_i + \Lambda_i +\frac{\Delta_i}{3}(4+\Omega_i-2\Lambda_i)
\approx \frac{5\Delta_i}{3}.
\ee

In fact, we typically obtain values of $\Delta_i\sim0.3$ in our study.
Keeping only linear terms in $\Delta_i$ results in errors of $\sim25$ per cent at $t=t_i$, and we expect them to grow as the system evolves.
We will thus depart from \citet{Ascasibar+04} and take the more accurate initial conditions
\be
\alpha(t_i)=(1+\Delta_i)^{1/3}
\ee
and compute $K_i$ by solving
\be
H_i t_i =
 \int_0^{(1+\Delta_i)^{-1/3}} \frac{ \dd\alpha }{ \sqrt{\Omega_i\alpha^{-1}+\Lambda_i\alpha^2-K_i} }
= \int_0^1         \frac{ \dd a }    { \sqrt{\Omega_i a^{-1}    +\Lambda_i a^2} },
\label{eqKi}
\ee
that is, by imposing the correct initial time for every shell
\footnote{For $\Omega_i\simeq1$ and $\Lambda_i\simeq0$, a good approximation at high redshift, the implicit equation (\ref{eqKi}) would reduce to $K_i^{-3/2}\int_0^{K_i(1+\Delta_i)^{-1/3}} \sqrt{\frac{x}{1-x}} \dd x = \frac{2}{3}$.}.
The initial velocity can be easily obtained as
\be
\dot\alpha(t_i) = \sqrt{ \Omega_i(1+\Delta_i)^{1/3} + \Lambda_i(1+\Delta_i)^{-2/3} - K_i }.
\ee

For positive overdensities, each shell will expand slower than the average universe, and, for high enough $\Delta_i$, will eventually turn around at some point $R_{\rm ta} \equiv r(T_{\rm ta})$ and start collapsing, crossing the inner shells on its way towards the centre.
After shell-crossing (which we approximate as immediately after turn-around), the appropriate integral of motion is
\be
\epsilon_2(x) =
\frac{\dot r^2}{2} +\phi(r) -G\frac{\frac{4\pi}{3}\rho_\Lambda r^3}{r} +\frac{j^2}{2r}
\ee
where $\phi(r,t)$ is the gravitational potential of the halo and $j$ is the average modulus of the specific angular momentum of the particles constituting the shell.
We approximate the local potential by a power-law mass distribution
\be
\phi(r)=\int_R^r \frac{GM(x)}{x^2} \dd x \approx
\int_R^r \frac{GM_x}{R} \left(\frac{x}{R^2}\right)^{\gamma-2} \dd x =
\frac{GM_x}{R} f(r/R)
\ee
where $R$ is the maximum radius of the orbit, $\gamma$ is the local logarithmic slope of the mass profile, $f(x)=\ln(x)$ for $\gamma=1$ and $f(x)=\frac{1}{\gamma-1}(x^{\gamma-1}-1)$ otherwise.
Concerning angular momentum, we adopt the prescription
\be
j^2 = \eta GMR
\label{eq:eta}
\ee
where $\eta$ is a free parameter from 0 (radial) to 1 (circular orbits).
For test particles orbiting a point mass ($\gamma=0$), the eccentricity of the orbits would be $e=1-\eta$, while the pericentric radius would be $r_{\rm min}/R=(1-e)/(1+e)=\eta/(2-\eta)$.
The equation of motion, written in terms of $\lambda\equiv r/R \equiv \alpha/\alpha_R$, is
\be
\dot \lambda^2 =
\Lambda_i (\lambda^2-1) +
\frac{\Omega_i}{\alpha_R^3} \left[ \frac{\eta}{2} (1-\lambda^{-2}) -f(\lambda) \right].
\label{eqMotion2}
\ee

Initially, $R=R_{\rm ta}$, but the potential $\phi(r)$ is not static.
Outer shells that collapse later will add a certain amount of mass, $M_{\rm add}(r)$, and the apocentre $R$ will slowly move inwards.
If $\phi(r)$ changes slowly compared to the orbital period, both the radial action
\be
J_r=\oint \dot r \dd r \propto \sqrt{R\, M(R)}
\ee
and the angular action (the specific angular momentum $j$) should be conserved.
We find the final apocentric radius, $R'$, by solving the implicit equation
\be
\frac{R'}{R}=\frac{M_x}{M_x+M_{\rm add}(R')}
\label{eqR'}
\ee
numerically.
Constant angular momentum implies $\eta'=\eta$, while the added mass should be taken into account by
\be
\frac{\Omega_i'}{\Omega_i}=\frac{R}{R'}.
\ee

In order to compute $M_{\rm add}(r)$, we assume instantaneous phase mixing.
After turn-around, each spherical shell turns into a density distribution with cumulative mass proportional to the fraction of time its constituent particles spend within $r$,
\be
\dd M_{\rm add}(r) = \frac{\dd M_x}{T} \int_0^{r/R} \frac{\dd\lambda}{\dot\lambda}
\label{eqdMadd}
\ee
where $dM_x$ is the mass of the shell and $T=\int_0^1 \dd\lambda/\dot\lambda$ its orbital period.
Neglecting $\Lambda_i$,
\be
M_{\rm add}(r)
\approx \int_{M_x}^{M_m} \tau\!\left(\!\frac{r}{R(M)}\!\right) \dd M
\ee
with
\be
\tau(x) \equiv
\frac{\int_0^x\, [\frac{\eta}{2}(1-\lambda^{-2})-f(\lambda)]^{-1/2}\,\dd\lambda }
     {\int_0^1\, [\frac{\eta}{2}(1-\lambda^{-2})-f(\lambda)]^{-1/2}\,\dd\lambda }.
\ee

\subsection{Numerical scheme}

First of all, the program computes the initial conditions from a given mass profile, $M(r)$, obtained from a simulation snapshot at redshift $z_{\rm IC}$:
\be
\Delta_i = \left( \frac{3M}{4\pi\Omega_m\rho_c r^3} -1 \right)(1+z_{\rm IC})a_i.
\ee
In our case, the snapshot corresponds to $z_{\rm IC}=50$.
The initial time for the SIM integration is set by the value of $a_i$, which we choose, for the sake of simplicity, to coincide with the time of the simulation snapshot, i.e. $a_i=1/(1+z_{\rm IC})$.

Then, the evolution of $N_{\rm s}$ logarithmically-spaced shells is integrated numerically, starting by the outermost, less overdense one.
For shells still expanding at the final epoch, $a_0$, one simply computes $\alpha_0=\alpha(t_0)$ from~(\ref{eqMotion1}).
The ensuing profile is, not surprisingly,
\be
M(xa_i\alpha_0)=M_x.
\ee

The situation becomes more complicated after turn-around, where
\be
M(R')=M_x + M_{\rm add}(R').
\ee
For these shells, we compute $R_{\rm ta}(M_x)=xa_i\alpha_{\rm ta}$ by finding the zero of equation~(\ref{eqMotion1}).
Then we compute $R'(M_x)$ according to~(\ref{eqR'}) and add the contribution of this shell to $M_{\rm add}(r)$
\footnote{$M_{\rm add}(r)$ is tabulated in a set of 5000 logarithmic bins and initially set to 0.}
using equation~(\ref{eqdMadd}).
In order to obtain a smooth transition between the two regimes, the contribution to $M_{\rm add}(r)$ of shells with turn-around time $T_{\rm ta}>t_0/2$ is multiplied by a factor $t_0/T_{\rm ta}-1$.

The process is repeated iteratively, decreasing the shell mass $M_x$ by $1$ per cent, until the initial radius is smaller than the innermost data point in the initial conditions.

\bibliography{SIM}

%%----------------
\end{document}